\documentclass[12pt,a4paper]{article}
\pdfoutput=1
\usepackage{color}
\usepackage{amssymb,amsmath,bm,bbold}
\usepackage{epsf}
\usepackage{epsfig}
\usepackage{relsize}
\usepackage[dvipsnames]{xcolor}
\usepackage[linktoc=page,bookmarks=false,colorlinks=false,linkbordercolor=RoyalBlue,citebordercolor=ForestGreen,urlbordercolor=CornflowerBlue]{hyperref}
\usepackage{latexsym,mathrsfs,dsfont}
\usepackage[normalem]{ulem} 
\usepackage[compress]{cite}
\usepackage{graphicx}
\usepackage{url}
\usepackage{booktabs}
\usepackage{float}
\usepackage{multirow}
\usepackage{changepage}
\usepackage[hypcap]{caption, subcaption}

\usepackage[titles]{tocloft}

\usepackage{tabularx,colortbl}

\usepackage{fancyhdr}
\advance \headheight by 3.0truept       

\allowdisplaybreaks[1]

\definecolor{red}{cmyk}{0,1,1,0.4}
\definecolor{darkgreen}{rgb}{0.0,0.6,0.0}
\definecolor{cDarkGrey}{RGB}{91,91,91}
\definecolor{cGrey}{RGB}{245,243,238}
\definecolor{cBlue}{RGB}{0,110,191}
\definecolor{cLightBlue}{RGB}{214,237,252}
\definecolor{cRed}{RGB}{196,0,100}
\definecolor{cLightRed}{RGB}{254,222,237}
\definecolor{cGreen}{RGB}{0,166,80}
\definecolor{cLightGreen}{RGB}{254,222,237}
\definecolor{cOrange}{RGB}{221,74,44}
\definecolor{cLightOrange}{RGB}{255,215,210}
\definecolor{cPurple}{RGB}{93,35,125}
\definecolor{cLightPurple}{RGB}{241,230,252}
\definecolor{cYellow}{RGB}{252,191,10}
\definecolor{cISSRBlue}{RGB}{0,111,174}
\definecolor{cISSRGrey}{RGB}{167,169,172}

\newcommand{\beq}{\begin{equation}}
\newcommand{\eeq}{\end{equation}}
\newcommand{\be}{\begin{equation}}
\newcommand{\ee}{\end{equation}}
\newcommand{\bi}{\begin{itemize}}
\newcommand{\ei}{\end{itemize}}
\newcommand{\ba}{\begin{array}}
\newcommand{\ea}{\end{array}}
\newcommand{\beqa}{\begin{eqnarray}}
\newcommand{\eeqa}{\end{eqnarray}}
\newcommand{\bea}{\begin{eqnarray}}
\newcommand{\eea}{\end{eqnarray}}
\newcommand{\beqn}{\begin{eqnarray}}
\newcommand{\eeqn}{\end{eqnarray}}

\newcounter{TODO}

\def \refsec#1{Section~\ref{#1}}

\def \reffig#1{Figure~\ref{#1}}

\def \reftab#1{Table~\ref{#1}}

\newcommand{\oL}[1]{\overline{#1}}

\newcommand{\wH}[1]{\widehat{#1}}

\newcommand{\ord}[1]{\mathcal{O}\left( #1 \right)}

\newcommand{\MSbar}{${\overline{\text{MS}}}$}

\newcommand{\cB}{\mathcal{B}}

\newcommand{\cL}{\mathcal{L}}

\newcommand{\cN}{\mathcal{N}}

\newcommand{\MeV}{\,\text{MeV}}

\newcommand{\GeV}{\,\text{GeV}}

\newcommand{\vcb}{|V_{cb}|}
\newcommand{\vtd}{|V_{td}|}

\newcommand{\vts}{|V_{ts}|}

\def\klpn{K_{L}\rightarrow\pi^0\nu\bar\nu}

\newcommand{\alS}{\alpha_s}

\newcommand{\alE}{\alpha_\text{em}}

\newcommand{\GF}{G_{\!F}}

\newcommand{\Lqcd}{\Lambda_\text{QCD}}

\newcommand{\muEW}{{\mu_\text{ew}}}

\newcommand\cdt[1]{\cdot10^{#1}}

\begin{document}


\vspace{-14mm}
\begin{flushright}
  AJB-21-2
\end{flushright}

\medskip

\begin{center}
{\large\bf\boldmath
  Searching for New Physics with 
  $\overline{\mathcal{B}}(B_{s,d}\to\mu\bar\mu)/\Delta M_{s,d}$
}
\\[1.0cm]
{\bf
  Christoph~Bobeth and 
  Andrzej~J.~Buras
}\\[0.3cm]

{\small
TUM Institute for Advanced Study,
    Lichtenbergstr. 2a, D-85747 Garching, Germany \\[0.2cm]
}
\end{center}

\vskip 0.5cm

\begin{abstract}
\noindent
We reemphasize that the ratio $R_{s\mu} \equiv \overline{\mathcal{B}}
(B_s\to\mu\bar\mu)/\Delta M_s$ is a measure of the tension of
the Standard Model (SM) with latest measurements of
$\overline{\mathcal{B}}(B_s\to\mu\bar\mu)$ that does not suffer
from the persistent puzzle on the $|V_{cb}|$ determinations from inclusive
versus exclusive $b\to c\ell\bar\nu$ decays and which affects the
value of the CKM element $\vts$ that is crucial for the SM predictions of
both $\overline{\mathcal{B}}(B_s\to\mu\bar\mu)$ and $\Delta M_s$, but
cancels out in the ratio $R_{s\mu}$.  
In our analysis we include higher order electroweak and QED corrections
and adapt the latest hadronic input to find a tension of about $2\sigma$
for $R_{s\mu}$ measurements with the SM independently of $\vts$.
We also discuss the ratio $R_{d\mu}$ which could turn out, in particular
in correlation with $R_{s\mu}$, to be useful for the search for New Physics,
when the data on both ratios improves. Also $R_{d\mu}$ is independent
of $\vcb$ or more precisely~$\vtd$.
\end{abstract}

\thispagestyle{empty}
\newpage
\setcounter{page}{1}

%
%
%
\section{Introduction}

Since the first observation of the decay $B_s \to \mu\bar\mu$ in 2013 there
have been steady improvements of the measurement of its branching ratio and
also of the one for $B_d \to \mu\bar\mu$ by CMS, LHCb and ATLAS collaborations
\cite{Sirunyan:2019xdu, Aaij:2017vad, Aaboud:2018mst}. In 2020 the three 
experimental collaborations combined their results to provide
the world average of the two-dimensional likelihood in the space of
$\oL{\cB}(B_s\to \mu\bar\mu)$ and ${\cB}(B_d\to \mu\bar\mu)$, which
give the one-dimensional results \cite{LHCb:2020zud, CMS:2020rox, ATLAS:2020acx}
\begin{align}
  \label{eq:LHCb2}
  \oL{\cB}(B_s \to \mu\bar\mu) & = (2.69^{+0.37}_{-0.35}) \cdt{-9}, 
\\
      \cB (B_d \to \mu\bar\mu) & < 1.6\, (1.9) \cdt{-10}
      \text{ at 90\% (95\%) CL} \,.
\end{align}

Very recently the LHCb collaboration presented their final results based
on the full Run-II data \cite{LHCb:2021awg, LHCb:2021vsc} 
\begin{align}
  \label{eq:LHCb2021}
  \oL{\cB}(B_s \to \mu\bar\mu) & 
  = \left( 3.09^{\,+0.46 \;+0.15}_{\,-0.43 \;-0.11} \right) \cdt{-9}, 
\\
      \cB (B_d \to \mu\bar\mu) & < 2.6 \cdt{-10}
      \text{ at 95\% CL}\,,
\end{align}
which imply new world averages. The world averages must be performed by the
experimental collaborations themselves to account properly for all systematic
uncertainties. Until then only provisional averages with varying sophistication
are available, as for example presented in \cite{Geng:2021nhg},
\cite{Altmannshofer:2021qrr} and \cite{Hurth:2021nsi}. We will use here
exemplary the value of \cite{Hurth:2021nsi}
\begin{align}
  \label{eq:WAV-Bs}
  \oL{\cB}(B_s \to \mu\bar\mu) & 
  = (2.85\; {}^{+0.34}_{-0.31}) \cdt{-9} \,,
\\
  \label{eq:WAV-Bd}
      \cB (B_d \to \mu\bar\mu) &
  <  2.05 \cdt{-10} \text{ at 95\% CL} \,.
\end{align}
The other preliminary world averages of $\oL{\cB}(B_s \to \mu\bar\mu)$ 
are given in \cite{Geng:2021nhg} and \cite{Altmannshofer:2021qrr} with
very similar values $(2.84\pm 0.33) \cdt{-9}$ and $(2.93\pm 0.35) \cdt{-9}$,
respectively. The upper bounds on $\cB (B_d \to \mu\bar\mu)$ are read off
from  the 2$\sigma$ contours of the 2-dimensional likelihood plots in
\cite{Hurth:2021nsi}, \cite{Geng:2021nhg} and \cite{Altmannshofer:2021qrr},
where the latter two find $2.0 \cdt{-10}$ and $2.2 \cdt{-10}$ as
upper bounds.

On the other hand the  present SM values of $\cB (B_q \to \mu\bar\mu)$, based on
the calculations over three decades by several groups \cite{Buchalla:1993bv,
Buchalla:1998ba, Buras:2012ru, Bobeth:2013uxa, Bobeth:2013tba, Hermann:2013kca,
Beneke:2017vpq, Beneke:2019slt}, read
\begin{align}
  \label{eq:LHCbTH}
  \oL{\cB}(B_s \to \mu\bar\mu)_\text{SM} &
  = (3.66 \pm 0.14) \cdt{-9} ,
\\
  \cB(B_d \to \mu\bar\mu)_\text{SM} &
  = (1.03 \pm 0.05) \cdt{-10} .
\end{align}
Comparing the results in \eqref{eq:WAV-Bs} with \eqref{eq:LHCbTH} implies the
tension between the SM and the data in the ballpark of $2\sigma$
\cite{Geng:2021nhg, Altmannshofer:2021qrr}.

We would like to point out that such a conclusion is premature because in
obtaining the result in \eqref{eq:LHCbTH} the {\em inclusive} determination
of $\vcb$ has been used with the value $\vcb_{B\to X_c} = (42.00 \pm 0.64)
\cdt{-3}$~\cite{Gambino:2016jkc}. For the corresponding
{\em exclusive}
determination of $\vcb$, as for example $\vcb_{B\to D} = (40.7 \pm 1.1) \cdt{-3}$
from $B\to D\ell\bar\nu$ \cite{Bordone:2019guc}, one finds the
branching ratio in question in the ballpark of $(3.44\pm 0.20) \cdt{-9}$
and the reduced tension of $1.4\sigma$ deeming the hopes for seeing new
physics in this decay at work. Full compatibility between theory and experiment
can be found with the less reliable determination $\vcb_{B\to D^*} = (38.8 \pm
1.4) \cdt{-3}$ from $B\to D^*\ell\bar\nu$ \cite{Bordone:2019guc},
which gives $\oL{\cB}(B_s \to \mu\bar\mu)_\mathrm{SM} = (3.12\pm 0.23) \cdt{-9}$.
Therefore, taking all these results into account, in our view the uncertainty
of $4\%$ in \eqref{eq:LHCbTH} does not represent properly the present
uncertainty in the SM prediction for the branching ratio in question. It is
significantly larger because of the $V_{cb}$ puzzle. We stress that it is 
only the parametrical CKM uncertainty. The remaining theoretical ones are
in the ballpark of a few percent.

In view of the fact that the tension between the inclusive and exclusive
determinations of $\vcb$ has not been satisfactorily resolved despite the
efforts of world experts lasting already for two decades (see
\cite{Gambino:2019sif} and references therein), it may still take a few years
before  we will be able to find out whether the tension between the data and the SM
value for the branching ratio in question is $2\sigma$ or significantly lower.

In this paper we would like to demonstrate that a much better insight in what
is going on can be obtained by using the strategy that one of us proposed already
in 2003 \cite{Buras:2003td}. In this strategy one considers instead of the
branching ratios the ratios
\begin{align}
  \label{eq:def:Rql}
  R_{q\mu} &
  \equiv \frac{ \oL{\cB}(B_q \to \mu\bar\mu)}{\Delta M_q}
&
  q & = d,s
\end{align}
that have the following advantages over the branching ratios themselves:
\begin{itemize}
\item
  The dependence on $\vcb$ drops out. Even more, the dependences on $\vts$
  and $\vtd$, that contain additional subleading uncertainties beyond $\vcb$
  cancel out.
\item
  The dependence on the $B_q$-meson decay constant $f_{B_q}$ drops out
  and present uncertainties in $f_{B_q}$ from lattice QCD (LQCD) 
  are irrelevant in this strategy.
\item
  The dependence on the top-quark mass is  decreased lowering thereby 
  the uncertainty due to $m_t$. 
\item
  Due to the negligible experimental errors on $\Delta M_q$, the experimental
  errors of $R_{q\mu}$ are practically the same as in the branching ratios
  themselves.
\end{itemize}
This means that for the purpose of testing the SM now the decision
of whether inclusive or exclusive value of $\vcb$ should be used is irrelevant
and as a byproduct the parametric uncertainty related to $f_{B_q}$ is absent
as well and the one due to $m_t$ is reduced.  However, it
should be emphasized that the main goal in the strategy of \cite{Buras:2003td}
is to test the SM and when $R_{s\mu}$ and $ R_{d\mu}$ are taken together
to test the models with Constrained Minimal Flavour Violation (CMFV) 
\cite{Buras:2000dm, Blanke:2006ig}. In this manner the possible anomalies
in the the ratios $R_{q\mu}$ would signal NP not only beyond the SM but also
beyond CMFV, that is non-SM operators and/or new flavour-violating parameters
beyond the CKM ones, in particular new CP-violating phases.

Yet nothing is for free. The use of $\Delta M_q$ introduces the dependence on
the non-perturbative parameters $B_q$ or $\wH B_q$. However, these parameters
are already known from LQCD and HQET sum rule calculations within a few
percent accuracy and the prospects for obtaining even better determinations
in coming years are good. We will be more explicit about it below.

At first sight one would think that the same result could be obtained in
global $b\to s\ell\bar\ell$ fits, that  use only $\Delta B = 1$ transitions,
by including now $\Delta M_s$. However, without the imposition of CMFV
and without a careful inclusion of the correlation between $\Delta M_s$
and $B_s\to\mu\bar\mu$ the cancellation of $\vcb$ in question can not
be achieved.

The outline of our paper is as follows. In \refsec{BF} we recall the
SM expressions for the two quantities from which $R_{q\mu}$ are
constructed and introduce a qualitatively similar ratio $\kappa_{q\mu}$.
In \refsec{NUM} we collect the numerical input and present the
numerical analysis of $R_{s\mu}$ and $\kappa_{s\mu}$. For completeness
we also present the result for $R_{d\mu}$ and $\kappa_{d\mu}$.
Further, we briefly discuss the double ratio $R_{s\mu}/R_{d\mu}$.
In \refsec{SUM} a brief outlook is given.

%
%
%
\section{Basic Formulae}
\label{BF}

In this section we recall the basic formulae for the branching ratios of the
leptonic decays $B_q\to\mu\bar\mu$ and the mass differences in neutral
$B$-meson systems $\Delta M_q$. Besides the higher order QCD corrections, we
include known next-to-leading (NLO) electroweak (EW) corrections as well as
QED corrections. 

The effective Lagrangian for $|\Delta B| = 1$ decays ($q=d,s$)
\begin{align}
  \label{eq:DB=1:eff:Lag}
  \cL_{\Delta B = 1} & 
  = \cN_q\, \sum_i C_i(\mu_b) \, O_i + \mbox{h.c.} \,, &
  \cN_q & 
    = \frac{\GF^2\, m_W^2}{\pi^2} V_{tb}^{} V_{tq}^\ast \,,
\end{align}
contains the normalization factor $\cN_q$, which is chosen to facilitate the
renormalization at NLO in EW interactions \cite{Misiak:2011bf, Bobeth:2013tba}.
The Wilson coefficients are evaluated at the scale $\mu_b \sim m_b$ of the
order of the $b$-quark mass and include NNLO QCD and NLO EW/QED corrections
\cite{Bobeth:2003at, Huber:2005ig, Bobeth:2013tba, Hermann:2013kca}.
At LO in EW/QED interactions the single operator
\begin{align}
  O_{10} & 
    = \big[\bar{q} \gamma^\mu P_L b \big] 
      \big[\bar\mu \gamma_\mu \gamma_5 \mu \big] \,, &
  P_L & \equiv \frac{1-\gamma_5}{2} ,
\end{align}
is relevant only.\footnote{Here we use the
  convention $C_{10} = - 2\, C_A$ compared to
  \cite{Hermann:2013kca, Bobeth:2013uxa} and $C_{10} = \widetilde{c}_{10}$ to
  \cite{Bobeth:2013tba}. It differs by a factor of sine-squared of the weak
  mixing angle to $c_{10}$ of \cite{Bobeth:2003at, Huber:2005ig}: $C_{10} = s_W^2
  c_{10}$ at LO in EW interactions.}  
The time-integrated branching fraction \cite{DeBruyn:2012wk}, denoted by a bar,
is given by
\begin{align}
  \label{eq:BR}
  \oL{\cB}(B_q \to \mu\bar\mu) &
  = \frac{|\cN_q|^2\, M_{B_q}^3\, f_{B_q}^2}{8 \pi\, \Gamma^H_q}
    \, \beta_{q\mu} \, \left(\frac{m_\mu}{M_{B_q}}\right)^2
    \Big|C_{10}^\text{eff}\Big|^2 ,
&
  \beta_{q\mu} & \equiv \sqrt{1 - \frac{4\, m_\mu^2}{M_{B_q}^2}} ,
\end{align}
where $C_{10}^\text{eff}$ includes
\begin{enumerate}
\item 
  the NLO EW corrections from matching the SM at the electroweak scale
  $\muEW \sim 160\GeV$ and resummed QED corrections to the scale $\mu_b
  \sim m_b$ \cite{Bobeth:2013tba}.
\item 
  power-enhanced structure-dependent NLO QED corrections between the
  scales $\mu_b$ and the scales $\Lqcd$ \cite{Beneke:2017vpq, Beneke:2019slt}.
\end{enumerate}
It is the photon-inclusive branching fraction, recovered after including 
soft-photon final-state radiation \cite{Buras:2012ru, Beneke:2019slt}.
In the SM the time-integration implies that the lifetime $\Gamma_q^H$ of
the heavy-mass eigenstate $|B_q^H\rangle$ has to be used instead of the
averaged one \cite{DeBruyn:2012wk}. However, time-integration is at the
current precision numerically only relevant for $B_s$ decays. We follow
\cite{Beneke:2019slt} for the calculation of $\oL{\cB}(B_q \to \mu\bar\mu)$.

The mass difference of the neutral meson system is governed in the SM
by a single $|\Delta B| = 2$ operator\footnote{This is also the
case of CMFV models, but the function $S_0(x_t)$ receives additional
flavour-universal contributions.}
\begin{align}
  \label{eq:DB=2:eff:Lag}
  \cL_{\Delta B = 2} & 
  = - \frac{\cN_q\,  V_{tb}^{} V_{tq}^\ast}{4}\;
    C_\text{VLL}(\mu_b) \, O_\text{VLL} + \text{h.c.} \,, 
&
  O_\text{VLL} &
  = \big[\bar q \gamma^\mu P_L b\big] \big[\bar q \gamma_\mu P_L b\big]
\end{align}
with
\begin{align}
  C_\text{VLL}(\muEW) &
  = S_0(x_t) + \ldots \,,
&
  S_0(x_t) &
  = \frac{4x_t-11x^2_t+x^3_t}{4(1-x_t)^2}
  - \frac{3x^3_t \ln x_t}{2(1-x_t)^3} ,
\end{align}
and $x_t \equiv m_t^2/m_W^2$.
Here we include besides the LO contribution $S_0(x_t)$ also higher order
corrections indicated by the dots. These are $S_\text{QCD}(x_t)$ at NLO in QCD 
\cite{Buras:1990fn} as well as $S_\text{ew}(x_t)$ at NLO in EW corrections
\cite{Gambino:1998rt}. The RG evolution from $\muEW$ to $\mu_b$ is performed
to NLO in QCD
\begin{align}
  \label{eq:C_VLL-mub}
  C_\text{VLL}(\mu_b) & =
  \eta^{6/23} \left[ S_0(x_t) + S_\text{ew}(x_t) +
  \frac{\alS(\mu_b)}{4\pi} \left(
   \frac{5165}{3174} (1 - \eta) S_0(x_t) +  \eta S_\text{QCD}(x_t) \right) \right] ,
\end{align} 
where $\eta\equiv \alS(\muEW)/\alS(\mu_b)$. The hadronic matrix element of
the $|\Delta B| = 2$ operator in the \MSbar{} scheme at the scale $\mu_b$
is defined by
\begin{align}
  \label{eq:def-DB2-me}
  \big\langle B_q \big| O_\text{VLL} \big| \oL{B}_q \big\rangle (\mu_b)
  = \frac{2}{3} M_{B_q}^2 f_{B_q}^2 B_q(\mu_b)
\end{align}
in terms of the $B_q$-meson decay constant $f_{B_q}$ and the so-called
bag factor $B_q(\mu_b)$ in the \MSbar{} scheme. The latter is related to the
renormalization group-invariant bag factor $\wH{B}_q$ at NLO in QCD as
\cite{Buras:1990fn}
\begin{align}
  \label{eq:conv-RGI-MSbar}
  \wH{B}_q & 
  =  \alS(\mu)^{-6/23}
    \left(1 + \frac{\alS(\mu)}{4\pi} \frac{5165}{3174} \right) B_q(\mu)
  \;\; \stackrel{\mu = 4.18\GeV}{=}\; 1.520 \, B_q(\mu = 4.18\GeV) ,
\end{align}
where $\alS(\mu = 4.18\GeV) = 0.2241$ has been used in the second equation.
Since the Wilson coefficient \eqref{eq:C_VLL-mub} is calculated in
the \MSbar{} scheme it is not advisable to convert lattice results that
were originally calculated in the \MSbar{} scheme to the RG-invariant bag
factor,\footnote{On the other hand if the matching between the lattice UV
regulator and the dimensional regulator is carried out non-perturbatively
one obtains directly $\wH B_q$ and one should not convert it to the
\MSbar{} scheme, because this would imply additional uncertainties. We 
thank Andreas Kronfeld for this insight.} as done for example by FLAG
before averaging them. At least the numerical values for such a conversion
should be always provided.
An even more principal question concerns the factorization of the matrix
element \eqref{eq:def-DB2-me} into the decay constant and the bag factor,
when lattice collaborations might actually calculate the l.h.s.
of \eqref{eq:def-DB2-me} directly. However, we stress that in our strategy
only the bag factor is required and our predictions profit from cancellations
of systematic uncertainties in lattice and sum rule predictions for this
quantity.

The mass difference reads as
\begin{align}
  \label{eq:Delta M_q}
  \Delta M_q &
  = \frac{G_{\!F}^2\, m_W^2}{6 \pi^2}  M_{B_q} f_{B_q}^2 \, B_q (\mu_b) \,
    |V_{tb}^{} V_{tq}^*|^2 \, \big|C_\text{VLL}(\mu_b)\big| \,.
\end{align}

The phenomenologically most interesting case is for $B_s$ mesons, since 
the leptonic decay has a larger branching ratio, enhanced by $(V_{ts}/V_{td})^2
\sim 20$ compared to $B_d$ mesons. In particular due to the unitarity of
the CKM matrix, the matrix element $V_{ts}$ depends strongly on the input
of $V_{cb}$ that should be preferably determined in tree-level decays
$b\to c\ell\bar\nu$. In fact the ratio
\begin{align}
  \label{eq:CKM-ratio}
  \left|\frac{V_{tb}^{} V_{ts}^*}{V_{cb}}\right|^2 &
  = 1- (1 - 2\rho) \lambda^2 +\ord{\lambda^4},
&
  \lambda & \approx 0.22,
\end{align}
with $\rho\approx 0.15$,
is rather independent of $B$-physics input, as can be seen in Wolfenstein
parametrization of the CKM matrix. This renders both, 
$\oL{\cB}(B_s \to \mu\bar\mu)$ and $\Delta M_s$, very sensitive to the
input value of $V_{cb}$. Unfortunately the persisting discrepancy of the
determination of $V_{cb}$ from inclusive and exclusive $b\to c \ell\bar\nu$ 
decays prevents stringent tests of the SM using charged-current tree-level
decays versus FCNC decays $\oL{\cB}(B_q \to \mu\bar\mu)$. Previous
predictions \cite{Bobeth:2013uxa, Beneke:2019slt} used the inclusive
determination of $|V_{cb}|_{B\to X_c}$, {because the theoretical predictions
are more solid for $B\to X_c\ell\bar\nu$, thereby yielding larger values} of
$\oL{\cB}(B_s \to \mu\bar\mu)$ compared to exclusive determinations. 
As pointed out in \cite{Buras:2003td} the dependence on $V_{cb}$
cancels out in the ratio $R_{s\mu}$, see \eqref{eq:def:Rql}, thus removing
the issue of $V_{cb}$ in tests of the SM with $\oL{\cB}(B_s \to \mu\bar\mu)$,
thereby introducing a correlation with $\Delta M_s$, which might involve
further assumptions on NP, like CMFV, when extending the tests beyond the
framework of the SM.

Analogous comments can be made about ${\cB}(B_d \to \mu\bar\mu)$ and
$\Delta M_d$ in which the CKM element $\vtd$ is involved. It is also
very sensitive to the value of $\vcb$ but it cancels out in the ratio
$R_{d\mu}$.

In summary the SM expression for $R_{q\mu}$ is given as follows:
\begin{equation}
  \label{eq:def:SMRq}
  R_{q\mu}|_\text{SM}   \;=\; 
  \frac{3(G_F \, m_W \, m_\mu)^2 \beta_{q\mu}} {4 \pi^3\Gamma^H_q}
  \frac{|C_{10}^\text{eff}|^2}{C_\text{VLL}(\mu_b)\, B_q(\mu_b)} \,.
\end{equation}

In addition to $R_{q\mu}$, that are dimensionful, slightly modified
dimensionless ratios
\begin{align}
  \label{eq:def:kappa}
  \kappa_{q\mu} &
  \;\equiv\; \frac{R_{q\mu} \, \Gamma^H_q}
         {(G_F \, m_W \, m_\mu)^2 \beta_{q\mu}} 
  \; \stackrel{\rm SM}{=} \;
  \frac{3 }{4 \pi^3}
  \frac{|C_{10}^\text{eff}|^2}{C_\text{VLL}(\mu_b)\, B_q(\mu_b)}
\end{align}
have been introduced in \cite{Bobeth:2013uxa}. The theory prediction for
$\kappa_{ql}$ does not suffer from the uncertainty of $\Gamma_q^H$, in
contrast to the theory prediction for $R_{q\mu}$. The uncertainty of
$\Gamma_q^H$ is now shifted to the experimental determination of
$\kappa_{q\mu}$. This could be an advantage, if the experimental
determination of the ratio $\Delta M_q/\Gamma_q^H$, which enters
$\kappa_{q\mu}$, allows for cancellation of systematic uncertainties
that would otherwise be present in $\Gamma_q^H$. However, currently
the uncertainty of the experimental determination of both, $R_{q\mu}$
and $\kappa_{q\mu}$, is dominated by the one of $\oL{\cB}(B_q \to
\mu\bar\mu)$. In the numerical analysis we will focus mainly on
$R_{q\mu}$. Further the overall dependence on $\GF$ in the SM prediction
of $R_{q\mu}$ is also removed in $\kappa_{ql}$. This would suggest
that its prediction is even independent of new physics contributions
in the $\beta$ decay $\mu\to e\, \nu_\mu \bar\nu_e$, but $\GF$ enters indirectly
in the determination of $m_W$ and the weak mixing angle when calculating
$C_{10}^\text{eff}$ and $C_\text{VLL}(\mu_b)$.

%
%
%
\section{Numerical Analysis}
\label{NUM}

The numerical predictions of the Wilson coefficients depend on the values
of the parameters of the SM from the electroweak sector, the strong coupling
$\alS$ and the top-quark mass $m_t$, which enter the calculation of the
Wilson coefficients. We collect their numerical values in \reftab{tab:num-input}
and proceed with the calculation of the Wilson coefficients as described
in \cite{Bobeth:2003at, Huber:2005ig, Bobeth:2013tba}. Note that we have chosen
for the input value of the top-quark mass in the pole-scheme the one determined
in cross-section measurements. For what concerns the electroweak renormalization,
we use the on-shell scheme 2 (``OS-2'') introduced in \cite{Bobeth:2013tba},
in which the mass of the $W$-boson is not an independent input, but calculated
following \cite{Awramik:2003rn}. Therefore the value in \reftab{tab:num-input}
differs slightly from the one in the PDG \cite{Zyla:2020zbs}. The central 
value of the matching scale is fixed to $\muEW = 160\GeV$ and the
central value of the low-energy scale is set to $\mu_b = 5.0\GeV$.

\begin{table}
\centering
\renewcommand{\arraystretch}{1.4}
\resizebox{\columnwidth}{!}{
\begin{tabular}{|lll|lll|}
\hline
  Parameter
& Value
& Ref.
&  Parameter
& Value
& Ref.
\\
\hline \hline
  $G_F$                & $1.166379\cdt{-5} \GeV^{-2}$ & \cite{Zyla:2020zbs}
& $m_Z$                & $91.1876(21)$ GeV      & \cite{Zyla:2020zbs}
\\
  $\alS^{(5)}(m_Z)$    & $0.1179(10)$           & \cite{Zyla:2020zbs}
& $m_W$                & $80.358(8)$ GeV        &
\\
  $\alE^{(5)}(m_Z)$    & $1/127.955$            & \cite{Tanabashi:2018oca}
& $m_t^\mathrm{OS}$    & $172.4(7)$ GeV         & \cite{Zyla:2020zbs}
\\
\hline
  $M_{B_s}$            & $5366.88(17)$ MeV      & \cite{Zyla:2020zbs}
& $M_{B_d}$            & $5279.65(12)$ MeV      & \cite{Zyla:2020zbs}
\\
  $\Delta M_s$         & $17.749(20)$ ps$^{-1}$ & \cite{Zyla:2020zbs}
& $\Delta M_d$         & $0.5065(19)$ ps$^{-1}$ & \cite{Zyla:2020zbs}
\\
  $1/\Gamma_s^H$       & $1.620(7)$ ps          & \cite{Zyla:2020zbs}
& $2/(\Gamma_d^H+\Gamma_d^L)$ & $1.519(4)$ ps   & \cite{Zyla:2020zbs}
\\
\hline
  $f_{B_s}$            & $230.3(1.3)$ MeV       & \cite{Aoki:2019cca}
& $f_{B_d}$            & $190.0(1.3)$ MeV       & \cite{Aoki:2019cca}
\\
  $B_s(4.18 \GeV)$     & $0.849(23)$            & \cite{DiLuzio:2019jyq}
& $B_d(4.18 \GeV)$     & $0.835(28)$            & \cite{DiLuzio:2019jyq}
\\
  $\wH B_s$            & $1.291(35)$            &
& $\wH B_d$            & $1.269(43)$            &
\\
  $\lambda_{B_s}(1\GeV)$ & $400(150)$ MeV       & \cite{Beneke:2020fot}
& $\lambda_{B_d}(1\GeV)$ & $350(150)$ MeV       & \cite{Beneke:2020fot}
\\
\hline
\end{tabular}
}
\renewcommand{\arraystretch}{1.0}
\caption{\label{tab:num-input}
  \small
  Numerical input values for parameters entering $\oL{\cB}(B_q \to \mu\bar\mu)$
  and $\Delta M_q$. The $B_q$-meson decay constants $f_{B_q}$ are averages from
  the FLAG group for $N_f = 2+1+1$ from \cite{Bazavov:2017lyh, Bussone:2016iua,
  Dowdall:2013tga, Hughes:2017spc}. They are almost identical to the single
  determination of FNAL/MILC $f_{B_s} = 230.7(1.3)\MeV$ and $f_{B_d} =
  190.5(1.3)\MeV$ \cite{Bazavov:2017lyh}. The bag factors have been
  converted from the \MSbar{} scheme to the RG-invariant form using the
  conversion factor 1.520 at $\mu = 4.18\GeV$ in \eqref{eq:conv-RGI-MSbar}.
}
\end{table}

The hadronic input would usually concern the $B$-meson decay constants $f_{B_q}$
and the bag factors $\wH B_q$ or $B_q(\mu_b)$, but in our procedure the
$f_{B_q}$ do not enter. We provide their values in \reftab{tab:num-input} 
for later purposes. The bag factors, on the other hand, are crucial
in this strategy and we summarize their present status below.

The FLAG averages of several $N_f =2+ 1$ lattice calculations
are \cite{Aoki:2019cca}
\begin{align}
  B_s(4.18 \GeV) & = 0.89(4) ,
&
  B_d(4.18 \GeV) & = 0.86(6) .
\end{align}
They are based\footnote{They have been converted to \MSbar{} using the
conversion factor 1.5158 from \cite{Bazavov:2016nty} in
\eqref{eq:conv-RGI-MSbar}.}
on the calculations \cite{Gamiz:2009ku, Aoki:2014nga,
Bazavov:2016nty} from HPQCD, RBC-UKQCD and MILC/FNAL, respectively.
The rather high values are driven by the calculation in \cite{Bazavov:2016nty}.
The more recent $N_f=2+1+1$ lattice calculation from HPQCD
\cite{Dowdall:2019bea}
\begin{align}
  B_s(4.16 \GeV) & = 0.813(35) ,
&
  B_d(4.16 \GeV) & = 0.806(40) ,
\end{align}
finds lower values and has smaller uncertainties. In particular they
provide an average with the $N_f=2+1$ results from MILC/FNAL 
\cite{Bazavov:2016nty}
\begin{align}
  B_s(4.16 \GeV) & = 0.84(3) ,
&
  B_d(4.16 \GeV) & = 0.83(3) .
\end{align}
Eventually HQET sum rule calculations of the bag factors are also available
\cite{King:2019lal}, which have been averaged in \cite{Lenz:2019lvd}. These
averages are based on \cite{Bazavov:2016nty, Dowdall:2019bea, King:2019lal}
and listed in \reftab{tab:num-input}. They will be used in the numerical
evaluations.

%
%
\subsection{\boldmath $R_{q\mu}$ and $\kappa_{q\mu}$}

The SM predictions for the ratios $R_{q\mu}$ are then
\begin{align}
  R_{s\mu}|_\text{SM} &
  = 2.042 \left(1 \;
     {}^{+0.0274}_{-0.0003} |_{\muEW} \;
     {}^{+0.0028}_{-0.0020} |_{\mu_b} \;
     {}^{+0.0101}_{-0.0100} |_{m_t} \;
     {}^{+0.0278}_{-0.0264} |_{B_s} \;
     {}^{+0.0043}_{-0.0043} |_{\Gamma_s^H}
  \right) \cdt{-10} \text{ ps}
\notag \\ &
  \label{eq:R_smu-SM}
  = \left(2.042 \; {}^{+0.083}_{-0.058} \right) \cdt{-10} \text{ ps} ,
\\[0.3cm]
  R_{d\mu}|_\text{SM} &
  = 1.947 \left(1 \;
     {}^{+0.0274}_{-0.0003} |_{\muEW} \;
     {}^{+0.0031}_{-0.0022} |_{\mu_b} \;
     {}^{+0.0101}_{-0.0100} |_{m_t} \;
     {}^{+0.0347}_{-0.0324} |_{B_d} \;
     {}^{+0.0026}_{-0.0026} |_{\Gamma_d}
  \right) \cdt{-10} \text{ ps}
\notag \\ &
  \label{eq:R_dmu-SM}
  = \left(1.947 \; {}^{+0.089}_{-0.066} \right) \cdt{-10} \text{ ps} .
\end{align}
They represent the most accurate predictions on these ratios to date.
Note that the central values would be $R_{s\mu}|_\text{SM} = 2.022 \cdt{-10}\,$ps
and $R_{d\mu}|_\text{SM} = 1.928 \cdt{-10}\,$ps when neglecting the NLO EW
corrections \cite{Gambino:1998rt} to $C_\text{VLL}$. The electroweak scale
has been varied within $\muEW \in [60,\, 300]\GeV$, and exhibits a strong
asymmetric effect, because the central value $\muEW = 160\GeV$ is close to
the lowest predictions of $R_{q\mu}$. The variation with $\muEW$ is rather
large, up to +3\%, mainly from $\muEW \to 60\GeV$. This simple variation
reproduces the more careful estimates of various higher order EW and QCD 
scheme dependences discussed for $\oL{\cB}(B_q \to \mu\bar\mu)$ in
\cite{Bobeth:2013uxa, Bobeth:2013tba}. The low-energy scale is varied within
$\mu_b \in [2.5,\, 10]\GeV$ and results in about 0.3\% uncertainty. The
top-quark mass dependence is about 1\% and the one of the lifetime about
$(0.3-0.4)\%$. The largest uncertainty of about $(3-4)$\%  is due to the
bag factors. The theoretical uncertainties of the observables have
been obtained by varying consecutively each parameter within the error
ranges given in \reftab{tab:num-input}. Throughout these uncertainties
are then combined by adding them in quadrature.

The experimental value of $R_{s\mu}$ follows from the preliminatry 
world average of $\oL{\cB}(B_s\to \mu\bar\mu)$ in \eqref{eq:WAV-Bs}
and $\Delta M_s$ given in \reftab{tab:num-input} as 
\begin{align}
  R_{s\mu}|_\text{exp} & =
  \left(1.61 \; {}^{+0.19}_{-0.17} \right) \cdt{-10} \text{ ps} \,,
\end{align}
having a tension with the SM prediction \eqref{eq:R_smu-SM} of about
$2.1\sigma$. Similar values $R_{s\mu} |_\text{exp} = (1.60 \pm 0.19)\cdt{-10}
\text{ps}$ and  $(1.65 \pm 0.20)\cdt{-10} \text{ps}$ are found from
the averages \cite{Geng:2021nhg} and \cite{Altmannshofer:2021qrr},
respectively, with tensions of $2.2\sigma$ and $1.9\sigma$. On the other
hand the preliminary world average of $\cB(B_d\to \mu\bar\mu)$ in
\eqref{eq:WAV-Bd} provides only an upper limit for
\begin{align}
  R_{d\mu}|_\text{exp} & < 4.05 \cdt{-10} \text{ ps}
  \text{ at 95\% CL} .
\end{align}

\begin{figure}
\centering
  \includegraphics[width=0.46\textwidth]{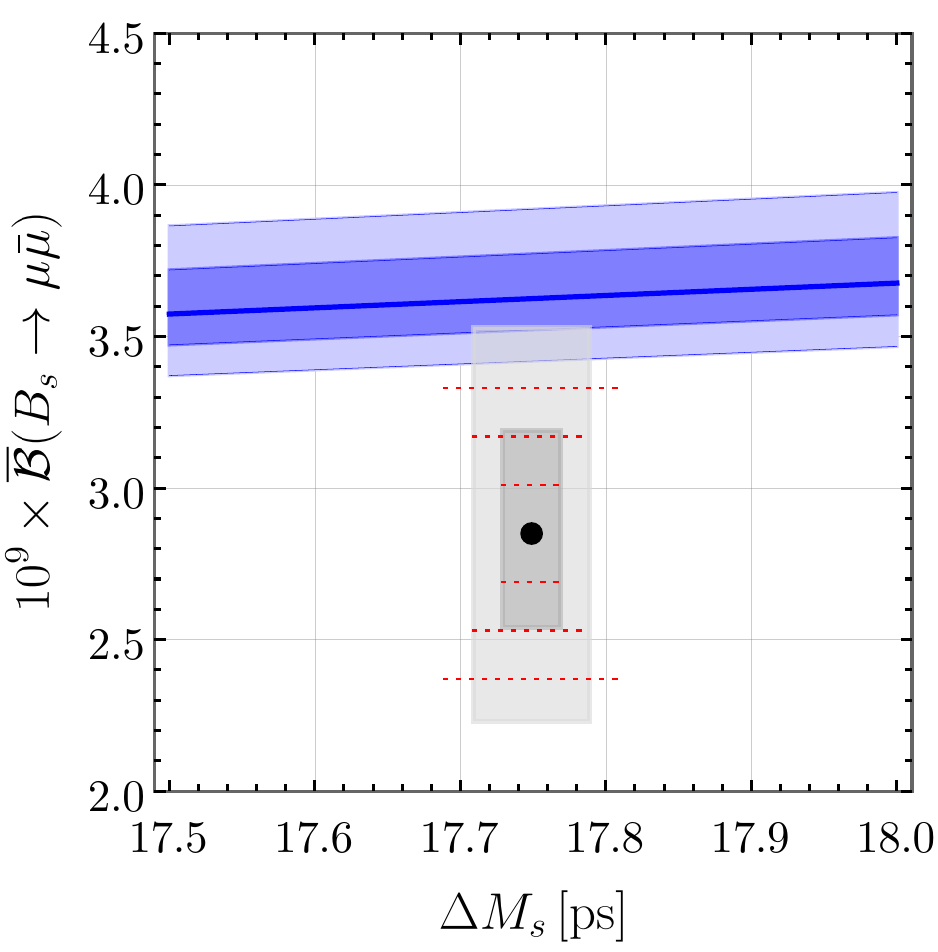}
  \hskip 0.05\textwidth
  \includegraphics[width=0.46\textwidth]{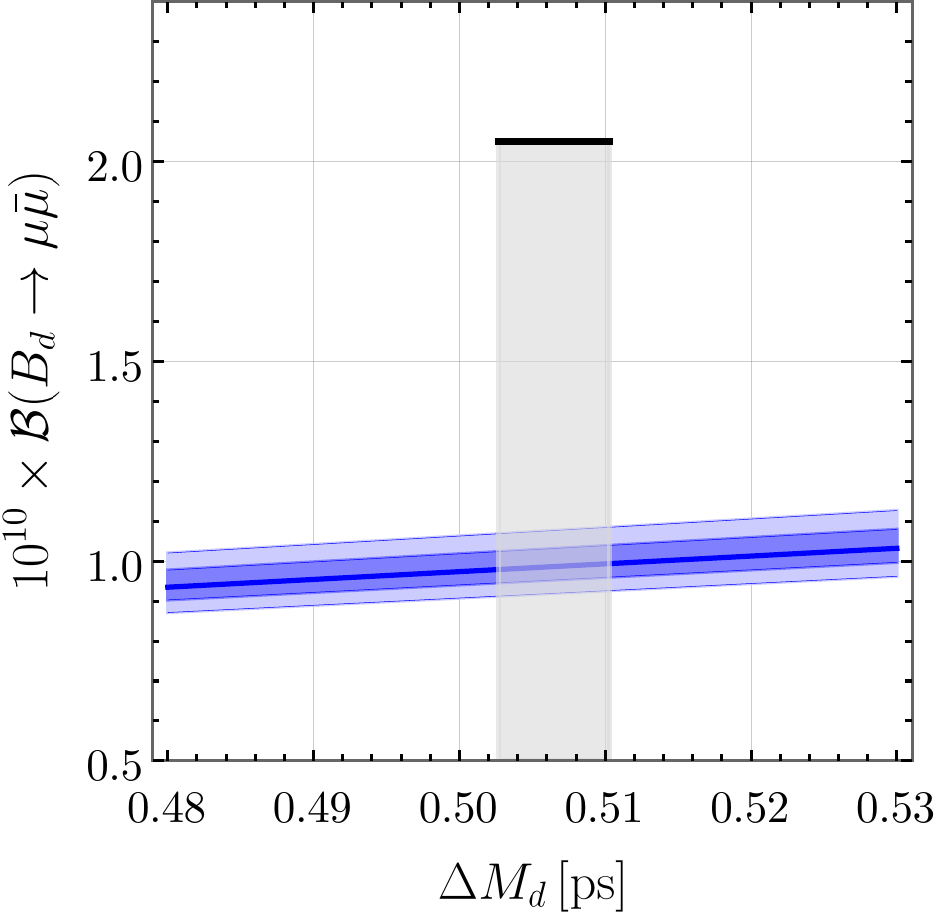}
\caption{\small The SM predictions of $R_{s\mu}$ [left] and $R_{d\mu}$ [right]
  are shown by the blue bands in the plane of $\oL{\cB}(B_q \to \mu\bar\mu)$
  versus $\Delta M_q$. The lighter bands indicate twice the theory errors
  given in \eqref{eq:R_smu-SM} and \eqref{eq:R_dmu-SM}, respectively.
  The (lighter) darker gray areas show the experimental results at (95\%) 
  68\%~CL from \cite{Hurth:2021nsi} with the dot at the central value
  and the solid line as upper bound,
  see \eqref{eq:WAV-Bs} and \eqref{eq:WAV-Bd}, respectively. The red dotted
  lines show the 68\%, 95\% and 99\% CL regions for the projections of LHCb
  with 300/fb, see text for more details, assuming the central value from 
  \cite{Altmannshofer:2021qrr}.
  }
  \label{fig:Br-vs-DelM}
\end{figure}

The SM predictions of $R_{q\mu}$ are compared in \reffig{fig:Br-vs-DelM}
with the experimental results in the plane of $\oL{\cB}(B_q \to \mu\bar\mu)$
versus $\Delta M_q$. In this plane the $R_{q\mu}$ are straight lines where
the bands indicate the theoretical uncertainties. The case of $R_{s\mu}$
shows a tension of about $2\sigma$, depending on the average used from 
\eqref{eq:WAV-Bs}. When interpreted in the context of physics beyond the
SM, the tension between experiment and SM in $R_{s\mu}$ could be caused
by new physics in both, $\Delta M_s$ and $\oL{\cB}(B_s\to\mu\bar\mu)$.
However, given the very tiny slope $R_{s\mu}|_\text{SM}$ the new physics
impact on $\Delta M_s$ should be rather large, i.e. the experimental
measurement should be around $14.5\,$ps to by fully compatible with
$R_{s\mu}|_\text{SM}$ and the measured value of $\oL{\cB}(B_s\to\mu\bar\mu)$.
The experimental prospects from LHCb with 300/fb for
the absolute uncertainty $\delta \oL{\cB}(B_s\to\mu\bar\mu) \approx 0.16
\cdt{-9}$ is about half the ones in \eqref{eq:WAV-Bs}, but it remains to
be seen whether future measurements confirm the current central values.
If so, the current tension would be increased to above $3\sigma$.

The SM predictions for the ratio $\kappa_{q\mu}$ are
\begin{align}
  \label{eq:SM-kappa}
  \kappa_{s\mu}|_\text{SM} &
  = \left(1.286 \; {}^{+0.052}_{-0.036} \right) \cdt{-2} ,
&
  \kappa_{d\mu}|_\text{SM} &
  = \left(1.308 \; {}^{+0.059}_{-0.044} \right) \cdt{-2} .
\end{align}
Note that the central values would be $\kappa_{s\mu}|_\text{SM} = 1.274 \cdt{-2}$
and $\kappa_{d\mu}|_\text{SM} = 1.295 \cdt{-2}$ when neglecting the NLO EW
corrections \cite{Gambino:1998rt} to $C_\text{VLL}$. The sources of uncertainty
are the same as for $R_{q\mu}$, except that the one of $\Gamma_q^H$ is removed.
For the remaining, the relative uncertainties in $\kappa_{q\mu}$ are the
same as in the corresponding $R_{q\mu}$.
The current experimental determinations are based on the world averages
\eqref{eq:WAV-Bs} and \eqref{eq:WAV-Bd}
\begin{align}
  \label{eq:WAV-kappa}
  \kappa_{s\mu}|_\text{exp} &
  = \left(1.011 \;{}^{+0.121}_{-0.110} \right) \cdt{-2},
&
  \kappa_{d\mu}|_\text{exp} &
  < 2.7 \cdt{-2} \text{ at 95\% CL}  ,
\end{align}
where the experimental error from $\Delta M_q$ and $\Gamma_q^H$ are negligible
at the current stage. Similar values $\kappa_{s\mu}|_\text{exp} =
(1.007 \pm 0.117) \cdt{-2}$  and $(1.039 \pm 0.124) \cdt{-2}$ are obtained
from \cite{Geng:2021nhg} and \cite{Altmannshofer:2021qrr}, respectively.
The corresponding upper bounds are $\kappa_{d\mu}|_\text{exp} <2.7 \cdt{-2}$
and $2.9 \cdt{-2}$, respectively.
The tension between the SM prediction and the experimental
measurement of $\kappa_{s\mu}$ is in the range $(1.8 - 2.2)\sigma$,
depending on the averages  presented in \cite{Geng:2021nhg, 
Altmannshofer:2021qrr, Hurth:2021nsi}.  There is only an upper bound for
$\kappa_{d\mu}|_\text{exp}$ well compatible with the SM prediction.

As pointed out in \cite{Buras:2003td}, the following relation holds
\begin{align}
  \label{eq:CMFV6}
    \frac{\Delta M_s}{\Delta M_d}
    \frac{\Gamma_d}{\Gamma_s^H}
    \frac{B_d(\mu_b)}{B_s(\mu_b)} &
  \; \stackrel{\text{CMFV}}{=} \; 
  \frac{\cB(B_s\to\mu\bar\mu)}{\cB(B_d \to\mu\bar\mu)} \,,
\end{align}
in the SM and also in any CMFV model, up to negligible effects. The l.h.s. of
\eqref{eq:CMFV6} involves only measurable quantities except for the ratio
$B_d/B_s$. This ratio can be determined with higher precision than the
individual bag factors. The most precise predictions are
$B_s/B_d = 0.9984(45)_\text{stat}(_{-63}^{+80})_\text{syst}$ \cite{Boyle:2018knm},
$B_s/B_d = 1.008(25)$ \cite{Dowdall:2019bea} and $B_s/B_d = 0.987(^{+7}_{-9})$
\cite{King:2019lal}. This leads to a relative uncertainty of about 
$(1-2)\%$ when using the values of the $\Delta M_q$ and the lifetimes from
\reftab{tab:num-input} together with the uncertainties of the ratio of bag
factors. Translating this result into ratios $R_{s\mu}$ and $R_{d\mu}$,
we find
\begin{align}
  \label{RR}
  \frac{R_{s\mu}}{R_{d\mu}} &
  \; \stackrel{\text{CMFV}}{=} \;
  \frac{\Gamma_d}{\Gamma_s^H}\frac{B_d(\mu_b)}{B_s(\mu_b)} 
  \;\; = \;\; \left\{ \begin{array}{ccc}
      1.072 \pm 0.011 & \qquad & \text{\cite{Boyle:2018knm}} \\[0.2cm]
      1.058 \pm 0.027 & \text{ for } & \text{\cite{Dowdall:2019bea}} \\[0.2cm]
      1.081 \pm 0.011 & \qquad & \text{\cite{King:2019lal}} 
    \end{array} \right. ,
\end{align}
a double ratio that is independent of CKM parameters and the Wilson
coefficients. The prospects to measure the ratio
$\cB(B_d\to\mu\bar\mu) /\oL{\cB}(B_s \to\mu\bar\mu)$ at LHCb foresee
a precision of 10\%~\cite{Bediaga:2018lhg} with 300/fb.

It should be emphasized that although the ratio of the two ratios in question
is common to all models with Constrained Minimal Flavour Violation (CMFV)
\cite{Buras:2000dm}, the ratios $R_{q\mu}$ themselves are not. Indeed
CMFV models can only be distinguished from each other by the Wilson coefficients
$C_{10}$ and $C_\text{VLL}$ entering \eqref{eq:BR} and \eqref{eq:Delta M_q},
respectively and varying them one just moves on the straight lines shown in 
\reffig{fig:Br-vs-DelM}.

Despite the comments just made, our result for the size of the tension in the
case of $R_{s\mu}$, that is independent of the value of $\vcb$, being in the
ballpark of $2\sigma$ is consistent with the results in \cite{Geng:2021nhg,
Altmannshofer:2021qrr}, where the inclusive value of $\vcb$ was used. But with
such a value the SM prediction for $\Delta M_s$ is fully consistent with the
data, although as analysed in \cite{DiLuzio:2019jyq}, with the improved future
theoretical calculations of $B_s$ and $f_{B_s}$ some amount of NP contributing
to $\Delta M_s$ could still be identified. Yet, these findings indicate that
$\Delta M_s$ is SM-like and it is some NP affecting $B_s\to\mu\bar\mu$ that is
dominantly responsible for the $2\sigma$ tension in $R_{s\mu}$ found by us.

%
%
\subsection{\boldmath $\Delta M_q$ and $V_{cb}$}

Assuming then for the moment that $\Delta M_q$ is SM-like,  we would
like to point out that the mass differences $\Delta M_q$
provide currently in the framework of the SM one of the most precise probes
of $|V_{tb}^{} V_{tq}^*|^2$, and hence indirectly also on $V_{cb}$. This is
thanks to the high experimental accuracy, but also to the high control over
the hadronic uncertainties from $f_{B_q}$ and $B_q(\mu)$ in the theoretical
predictions. In fact $\Delta M_q$ are presently the only loop induced
transitions in the SM in which both theoretical calculations and experimental
data are very accurate, even better than $B\to X_s\gamma$ decay and
$\varepsilon_K$, both known at the NNLO level. In principle also
$K^+\to \pi^+ \nu\bar\nu$ and $\klpn$, being theoretically clean 
\cite{Buras:2005gr, Buras:2006gb, Brod:2010hi}
and sensitive to the choice of $\vcb$, could be used for this purpose
\cite{Buchalla:1996fp},\footnote{Note that the $K\to\pi\nu\bar\nu$ branching
ratios being proportional to $|V_{td}^{} V_{ts}^*|^2$ are even more sensitive
to the choice of $\vcb$ than the $B$ observables considered here.} but this
would require dramatic improvements on the experimental side and from the
present perspective it is better to use them for the search of NP rather
than the determination of the CKM parameters.

The various sources of uncertainties in $\Delta M_q$ contribute as
\begin{align}
  \Delta M_s |_\text{SM} &
  = 10 444.8 \times |V_{tb}^{} V_{ts}^*|^2 \, \big( 1 \;
     {}^{+0.0001}_{-0.0269} |_{\muEW} \;
     {}^{+0.0008}_{-0.0009} |_{\mu_b} \;
     {}^{+0.0098}_{-0.0098} |_{m_t} \;
     {}^{+0.0271}_{-0.0271} |_{B_s} \;
     {}^{+0.0113}_{-0.0113} |_{f_{B_s}} 
  \big) \, \text{ps}^{-1}
\notag \\ &
  = 10 444.8 \times |V_{tb}^{} V_{ts}^*|^2 \,
    \big( 1 \; {}^{+0.031}_{-0.041} \big) \, \text{ps}^{-1} ,
\\[0.3cm]
  \Delta M_d |_\text{SM} &
  = \;\, 6878.3 \times |V_{tb}^{} V_{td}^*|^2 \, \big( 1 \;
     {}^{+0.0001}_{-0.0269} |_{\muEW} \;
     {}^{+0.0008}_{-0.0009} |_{\mu_b} \;
     {}^{+0.0098}_{-0.0098} |_{m_t} \;
     {}^{+0.0335}_{-0.0335} |_{B_d} \;
     {}^{+0.0137}_{-0.0137} |_{f_{B_d}} 
  \big) \, \text{ps}^{-1}
\notag \\ &
  = \;\, 6878.3 \times |V_{tb}^{} V_{td}^*|^2 \,
    \big( 1 \; {}^{+0.038}_{-0.046} \big) \, \text{ps}^{-1} ,
\end{align}
where the CKM combinations are left unspecified. The SM predictions
have about $(4-5)\%$ relative uncertainty, with the largest uncertainty
from the bag factor. This allows to extract the CKM combinations with
about 2\% relative uncertainty, which is at the same level
as the determination from the inclusive $B\to X_c \ell\bar\nu$ with
about $1.5\%$ relative uncertainty:
$\vcb_{B\to X_c} = (42.00 \pm 0.64) \cdt{-3}$~\cite{Gambino:2016jkc}.
The experimental measurements of $\Delta M_q$ yield the central values
$|V_{tb}^{} V_{ts}^*| = 41.22 \cdt{-3}$ and $|V_{tb}^{} V_{td}^*| =
8.58 \cdt{-3}$ in the SM. On the basis of the inclusive determination
of $\vcb_{B\to X_c}$ one finds in the framework of the SM for the ratio
in \eqref{eq:CKM-ratio} that $|V_{tb}^{} V_{ts}^*|/\vcb = 0.982$. It 
would be interesting to verify whether CKM fits that do not include $|\Delta B| = 2$
and $b\to c \ell\bar\nu$ processes provide values that are compatible
with this one. That the determinations of $|V_{tb}^{} V_{ts}^*|$ in
the framework of the SM from $\Delta M_s$ lead to branching ratios of
$\oL{\cB}(B_s\to \mu\bar\mu)$ above the data \eqref{eq:WAV-Bs} has been
discussed previously, as for example in \cite{Dowdall:2019bea,
DiLuzio:2019jyq}.

%
%
\subsection{\boldmath The issue of $m_t$ in rare decays}

\begin{figure}
\centering
  \includegraphics[width=0.46\textwidth]{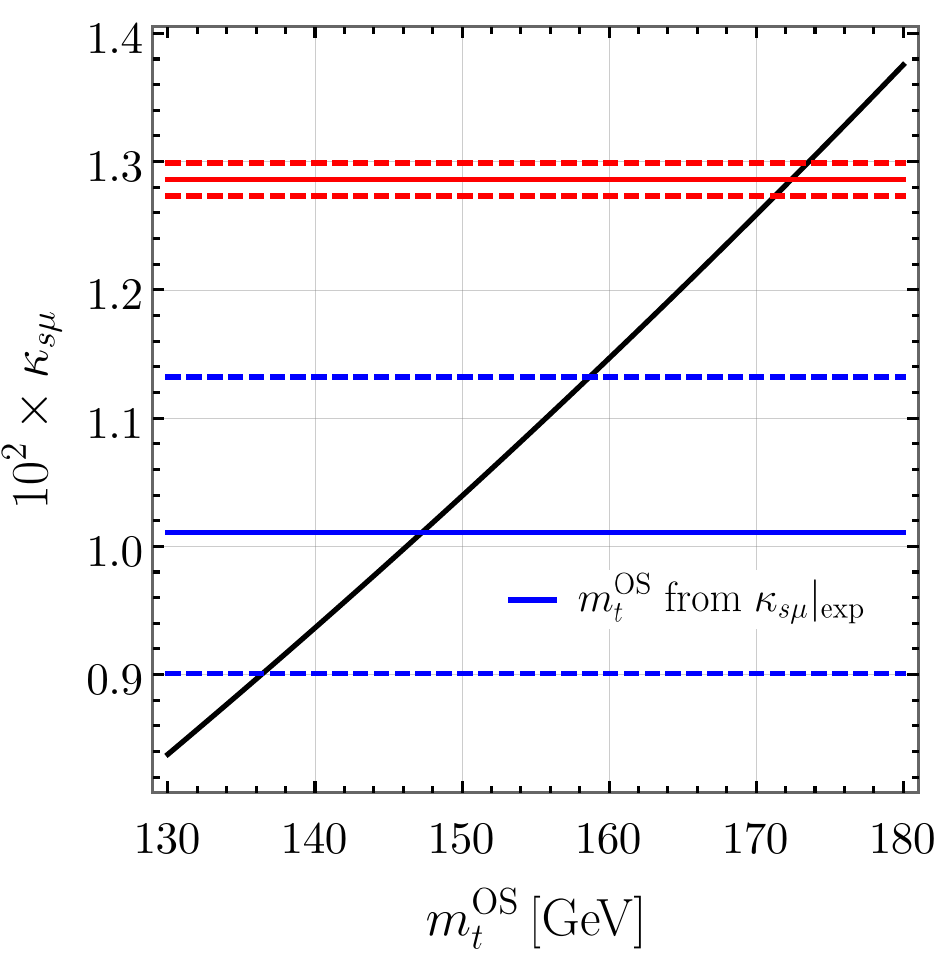}
\caption{\small The SM prediction of $\kappa_{s\mu}$ depending on the
  value of $m_t^\text{OS}$ [black]. The SM prediction in \eqref{eq:SM-kappa}
  [red] is for the value given in \reftab{tab:num-input} from top-cross-section
  determinations, showing the uncertainty of about 1\% on $\kappa_{s\mu}$
  [red dashed lines]. The experimental value $\kappa_{s\mu}|_\text{exp}$
  in \eqref{eq:WAV-kappa} [blue] requires much lower values $m_t^\text{OS}$.
}
  \label{fig:kappa-vs-mt}
\end{figure}

The ratios $\kappa_{q\mu} \sim (m_t^\text{\MSbar{}}/m_W)^2$ scale with the
second power of the top-quark mass in the \MSbar{} scheme and in principle
might be used to determine the top-quark mass in rare flavour processes under
the assumption that $\kappa_{q\mu}$ are not affected by NP contributions, which
is still a possibility. A discussion of other examples in flavour physics that
require the knowledge of CKM input and the corresponding prospects can be found
in \cite{Giudice:2015toa}. The \MSbar{} mass is actually the preferred scheme
for rare decay calculations over the
pole scheme (OS), which however is used in many collider physics applications
and provided in the PDG \cite{Zyla:2020zbs}. In the numerical evaluation we
converted the top-quark mass from the pole to the \MSbar{} scheme (in QCD),
see \cite{Bobeth:2013uxa}, using the peturbative expressions at 3-loops.
For illustration
we show the dependence of $\kappa_{s\mu}$ on  $m_t^\text{OS}$ in 
\reffig{fig:kappa-vs-mt}.  The preferred values of $m_t^\text{OS}$, corresponding
to the central value and 68\% CL interval of $\kappa_{s\mu}|_\text{exp}$ in
\eqref{eq:WAV-kappa}, are $m_t^\text{OS} \in [136,\, 158]\GeV$ and are much lower
than those from collider determinations. Such low values would correspond to
absolute stability
of the SM vacuum \cite{Buttazzo:2013uya}. The current experimental uncertainty
of $\kappa_{s\mu}|_\text{exp}$ is dominated by the one of $\oL{\cB}(B_s \to
\mu\bar\mu)$. Assuming a future measurement with 4\% relative uncertainty,
as might be feasible at LHCb, a determination of $m_t^\text{OS}$ with about
2\% relative uncertainty can be expected, if the theoretical uncertainties
due to the bag factor will be negligible at this level in the future.
Clearly this is not competitive to determinations based on collider observables.
If the $V_{cb}$ puzzle
will be solved in the future, then $\oL{\cB}(B_s \to \mu\bar\mu) \sim
(m_t^\text{\MSbar{}}/m_W)^4$ could offer a better opportunity, because
it scales with the fourth power of $m_t$, if sufficient control over
$V_{cb}$ and $f_{B_s}$ are provided.

%
%
%
\section{Summary and Outlook}
\label{SUM}

It is evident that once both ratios $R_{s\mu}$ and $R_{d\mu}$ will be measured
precisely the strategy presented in \cite{Buras:2003td} and executed here 18 years
later will provide one of the theoretically cleanest tests of the SM and more
generally of CMFV models.

Of course a global analysis of a large set of processes in the framework
of a specific model, as done in the standard analyses of the unitarity
triangle or the recent global fits testing the violation of lepton flavour
universality, will reveal any potential tension by lowering the
goodness of the fit. However, a clear-cut insight on the origin of tensions 
is not so easily obtainable. We emphasize that in contrast our proposed
strategy involves only two observables on the experimental side and requires
only well-accessible parametric input on the theoretical side, providing
optimal control and a very transparent test of the SM (and CMFV models).
Indeed, the global analyses involve usually CKM uncertainties, in
particular the one from $\vcb$, and also hadronic uncertainties present
in other processes that are larger than the ones in $\Delta M_s$ and
$B_s\to\mu\bar\mu$. Moreover,
NP could enter many observables used in such global fits and the transparent 
identification of the impact of NP on a given observable is a challenge.
On the contrary in the proposed ratios all these uncertainties cancel out
except for the bag factors, which are already precisely known from LQCD
and importantly do not depend on NP. In this manner concentrating just on 
$\Delta M_s$ and $B_s\to\mu\bar\mu$ allows us to test the SM (and CMFV models)
independently of the values of CKM parameters and also independently of whether
NP affects other observables or not. These ratios could turn out to be
smoking guns of new physics.

However, one should emphasize that taking ratios of observables cancels not
only parametric, theoretical and experimental uncertainties. It can in
principle cancel also NP effects present in our case in the two branching
ratios and in mass differences $\Delta M_{s,d}$. Therefore the complete search
for NP must also  consider four observables separately that brings back CKM
uncertainties. Yet, the analysis presented here allows to conclude, without
any use of the CKM parameters and the decay constants $f_{B_q}$, that indeed
new experimental results for $B_s\to \mu\bar\mu$ exhibit some footprints
of NP that affect the SM correlation between $\oL{\cB}(B_s\to \mu\bar\mu)$
and $\Delta M_s$. We are looking forward to improved results for
$B_s\to \mu\bar\mu$ and even more to improved results for $B_d\to \mu\bar\mu$
which would allow to test the correlation between $R_{s\mu}$ and $R_{d\mu}$
that as seen in \eqref{RR} is already precisely known within CMFV models.

\vskip 0.2cm

{\bf Acknowledgements}

\noindent
We would like to thank Andreas Kronfeld and Alexander Lenz for informative
discussions on the $B_q$ bag factors determined by means of LQCD and HEFT sum rules.
A.J.B acknowledges financial support from the Excellence Cluster ORIGINS,
funded by the Deutsche Forschungsgemeinschaft (DFG, German Research Foundation), 
Excellence Strategy, EXC-2094, 390783311.

%
%
%
\renewcommand{\refname}{R\lowercase{eferences}}

\addcontentsline{toc}{section}{References}

\bibliographystyle{JHEP}

\small

\bibliography{Bookallrefs}

\providecommand{\href}[2]{#2}\begingroup\raggedright\begin{thebibliography}{10}

\bibitem{Sirunyan:2019xdu}
{\bf CMS} Collaboration, A.~M. Sirunyan et~al., {\it {Measurement of properties
  of B$^0_\mathrm{s}\to\mu^+\mu^-$ decays and search for B$^0\to\mu^+\mu^-$
  with the CMS experiment}},  {\em JHEP} {\bf 04} (2020) 188,
  [\href{http://arxiv.org/abs/1910.12127}{{\tt arXiv:1910.12127}}].

\bibitem{Aaij:2017vad}
{\bf LHCb} Collaboration, R.~Aaij et~al., {\it {Measurement of the
  $B^0_s\to\mu^+\mu^-$ branching fraction and effective lifetime and search for
  $B^0\to\mu^+\mu^-$ decays}},  {\em Phys. Rev. Lett.} {\bf 118} (2017), no.~19
  191801, [\href{http://arxiv.org/abs/1703.05747}{{\tt arXiv:1703.05747}}].

\bibitem{Aaboud:2018mst}
{\bf ATLAS} Collaboration, M.~Aaboud et~al., {\it {Study of the rare decays of
  $B^0_s$ and $B^0$ mesons into muon pairs using data collected during 2015 and
  2016 with the ATLAS detector}},  {\em JHEP} {\bf 04} (2019) 098,
  [\href{http://arxiv.org/abs/1812.03017}{{\tt arXiv:1812.03017}}].

\bibitem{LHCb:2020zud}
{\bf LHCb} Collaboration, {\it {Combination of the ATLAS, CMS and LHCb results
  on the $B^0_{(s)} \to \mu^+ \mu^-$ decays}},
  \href{http://arxiv.org/abs/LHCb-CONF-2020-002}{{\tt LHCb-CONF-2020-002}}.

\bibitem{CMS:2020rox}
{\bf CMS} Collaboration, {\it {Combination of the ATLAS, CMS and LHCb results
  on the $B^0_{(s)} \to \mu^+\mu^-$ decays}},
  \href{http://arxiv.org/abs/CMS-PAS-BPH-20-003}{{\tt CMS-PAS-BPH-20-003}}.

\bibitem{ATLAS:2020acx}
{\bf ATLAS} Collaboration, {\it {Combination of the ATLAS, CMS and LHCb results
  on the $B^0_{(s)}\to\mu^+\mu^-$ decays.}},
  \href{http://arxiv.org/abs/ATLAS-CONF-2020-049}{{\tt ATLAS-CONF-2020-049}}.

\bibitem{LHCb:2021awg}
{\bf LHCb} Collaboration, R.~Aaij et~al., {\it {Measurement of the
  $B^0_s\to\mu^+\mu^-$ decay properties and search for the $B^0\to\mu^+\mu^-$
  and $B^0_s\to\mu^+\mu^-\gamma$ decays}},
  \href{http://arxiv.org/abs/2108.09283}{{\tt arXiv:2108.09283}}.

\bibitem{LHCb:2021vsc}
{\bf LHCb} Collaboration, R.~Aaij et~al., {\it {Analysis of neutral $B$-meson
  decays into two muons}},  \href{http://arxiv.org/abs/2108.09284}{{\tt
  arXiv:2108.09284}}.

\bibitem{Geng:2021nhg}
L.-S. Geng, B.~Grinstein, S.~J\"ager, S.-Y. Li, J.~Martin~Camalich, and R.-X.
  Shi, {\it {Implications of new evidence for lepton-universality violation in
  $b\to s\ell^+\ell^-$ decays}},  \href{http://arxiv.org/abs/2103.12738}{{\tt
  arXiv:2103.12738}}.

\bibitem{Altmannshofer:2021qrr}
W.~Altmannshofer and P.~Stangl, {\it {New Physics in Rare $B$ Decays after
  Moriond 2021}},  \href{http://arxiv.org/abs/2103.13370}{{\tt
  arXiv:2103.13370}}.

\bibitem{Hurth:2021nsi}
T.~Hurth, F.~Mahmoudi, D.~M. Santos, and S.~Neshatpour, {\it {More Indications
  for Lepton Nonuniversality in $b \to s \ell^+ \ell^-$}},
  \href{http://arxiv.org/abs/2104.10058}{{\tt arXiv:2104.10058}}.

\bibitem{Buchalla:1993bv}
G.~Buchalla and A.~J. Buras, {\it {QCD corrections to rare $K$ and $B$ decays
  for arbitrary top quark mass}},  {\em Nucl.~Phys.} {\bf B400} (1993)
  225--239.

\bibitem{Buchalla:1998ba}
G.~Buchalla and A.~J. Buras, {\it {The rare decays $K\to\pi \nu\bar\nu$, $B\to
  X \nu\bar\nu$ and $B\to \ell^+\ell^-$: An Update}},  {\em Nucl.~Phys.} {\bf
  B548} (1999) 309--327, [\href{http://arxiv.org/abs/hep-ph/9901288}{{\tt
  hep-ph/9901288}}].

\bibitem{Buras:2012ru}
A.~J. Buras, J.~Girrbach, D.~Guadagnoli, and G.~Isidori, {\it {On the Standard
  Model prediction for $\mathcal{B}(B_{s,d} \to \mu^+\mu^-)$}},  {\em
  Eur.~Phys.~J.} {\bf C72} (2012) 2172,
  [\href{http://arxiv.org/abs/1208.0934}{{\tt arXiv:1208.0934}}].

\bibitem{Bobeth:2013uxa}
C.~Bobeth, M.~Gorbahn, T.~Hermann, M.~Misiak, E.~Stamou, et~al., {\it
  {$B_{s,d}\to \ell^+ \ell^-$ in the Standard Model with Reduced Theoretical
  Uncertainty}},  {\em Phys.~Rev.~Lett.} {\bf 112} (2014) 101801,
  [\href{http://arxiv.org/abs/1311.0903}{{\tt arXiv:1311.0903}}].

\bibitem{Bobeth:2013tba}
C.~Bobeth, M.~Gorbahn, and E.~Stamou, {\it {Electroweak Corrections to $B_{s,d}
  \to \ell^+ \ell^-$}},  {\em Phys.~Rev.} {\bf D89} (2014) 034023,
  [\href{http://arxiv.org/abs/1311.1348}{{\tt arXiv:1311.1348}}].

\bibitem{Hermann:2013kca}
T.~Hermann, M.~Misiak, and M.~Steinhauser, {\it {Three-loop QCD corrections to
  $B_s \to \mu^+ \mu^-$}},  {\em JHEP} {\bf 1312} (2013) 097,
  [\href{http://arxiv.org/abs/1311.1347}{{\tt arXiv:1311.1347}}].

\bibitem{Beneke:2017vpq}
M.~Beneke, C.~Bobeth, and R.~Szafron, {\it {Enhanced electromagnetic correction
  to the rare $B$-meson decay $B_{s,d} \to \mu^+ \mu^-$}},  {\em Phys. Rev.
  Lett.} {\bf 120} (2018), no.~1 011801,
  [\href{http://arxiv.org/abs/1708.09152}{{\tt arXiv:1708.09152}}].

\bibitem{Beneke:2019slt}
M.~Beneke, C.~Bobeth, and R.~Szafron, {\it {Power-enhanced leading-logarithmic
  QED corrections to $B_q \to \mu^+\mu^-$}},  {\em JHEP} {\bf 10} (2019) 232,
  [\href{http://arxiv.org/abs/1908.07011}{{\tt arXiv:1908.07011}}].

\bibitem{Gambino:2016jkc}
P.~Gambino, K.~J. Healey, and S.~Turczyk, {\it {Taming the higher power
  corrections in semileptonic B decays}},  {\em Phys. Lett. B} {\bf 763} (2016)
  60--65, [\href{http://arxiv.org/abs/1606.06174}{{\tt arXiv:1606.06174}}].

\bibitem{Bordone:2019guc}
M.~Bordone, N.~Gubernari, D.~van Dyk, and M.~Jung, {\it {Heavy-Quark expansion
  for ${{\bar{B}}_s\rightarrow D^{(*)}_s}$ form factors and unitarity bounds
  beyond the ${SU(3)_F}$ limit}},  {\em Eur. Phys. J. C} {\bf 80} (2020), no.~4
  347, [\href{http://arxiv.org/abs/1912.09335}{{\tt arXiv:1912.09335}}].

\bibitem{Gambino:2019sif}
P.~Gambino, M.~Jung, and S.~Schacht, {\it {The $V_{cb}$ puzzle: An update}},
  {\em Phys. Lett. B} {\bf 795} (2019) 386--390,
  [\href{http://arxiv.org/abs/1905.08209}{{\tt arXiv:1905.08209}}].

\bibitem{Buras:2003td}
A.~J. Buras, {\it {Relations between $\Delta M_{s,d}$ and $B_{s,d} \to \mu^+
  \mu^-$ in models with minimal flavour violation}},  {\em Phys.~Lett.} {\bf
  B566} (2003) 115--119, [\href{http://arxiv.org/abs/hep-ph/0303060}{{\tt
  hep-ph/0303060}}].

\bibitem{Buras:2000dm}
A.~J. Buras, P.~Gambino, M.~Gorbahn, S.~J{\"a}ger, and L.~Silvestrini, {\it
  {Universal unitarity triangle and physics beyond the standard model}},  {\em
  Phys.~Lett.} {\bf B500} (2001) 161--167,
  [\href{http://arxiv.org/abs/hep-ph/0007085}{{\tt hep-ph/0007085}}].

\bibitem{Blanke:2006ig}
M.~Blanke, A.~J. Buras, D.~Guadagnoli, and C.~Tarantino, {\it {Minimal Flavour
  Violation Waiting for Precise Measurements of $\Delta M_s$, $S_{\psi \phi}$,
  $A^s_\text{SL}$, $|V_{ub}|$, $\gamma$ and $B^0_{s,d} \to \mu^+ \mu^-$}},
  {\em JHEP} {\bf 10} (2006) 003,
  [\href{http://arxiv.org/abs/hep-ph/0604057}{{\tt hep-ph/0604057}}].

\bibitem{Misiak:2011bf}
M.~Misiak, {\it {Rare B-Meson Decays}},  in {\em {Proceedings, 15th Lomonosov
  Conference on Elementary Particle Physics (LomCon): Particle Physics at the
  Tercentenary of Mikhail Lomonosov}}, pp.~301--305, 2013.
\newblock \href{http://arxiv.org/abs/1112.5978}{{\tt arXiv:1112.5978}}.

\bibitem{Bobeth:2003at}
C.~Bobeth, P.~Gambino, M.~Gorbahn, and U.~Haisch, {\it {Complete NNLO QCD
  analysis of $\bar B \to X_s \ell^+ \ell^-$ and higher order electroweak
  effects}},  {\em JHEP} {\bf 04} (2004) 071,
  [\href{http://arxiv.org/abs/hep-ph/0312090}{{\tt hep-ph/0312090}}].

\bibitem{Huber:2005ig}
T.~Huber, E.~Lunghi, M.~Misiak, and D.~Wyler, {\it {Electromagnetic logarithms
  in $\bar B\to X(s) l^+ l^-$}},  {\em Nucl.~Phys.} {\bf B740} (2006) 105--137,
  [\href{http://arxiv.org/abs/hep-ph/0512066}{{\tt hep-ph/0512066}}].

\bibitem{DeBruyn:2012wk}
K.~De~Bruyn, R.~Fleischer, R.~Knegjens, P.~Koppenburg, M.~Merk, et~al., {\it
  {Probing New Physics via the $B^0_s\to \mu^+\mu^-$ Effective Lifetime}},
  {\em Phys.~Rev.~Lett.} {\bf 109} (2012) 041801,
  [\href{http://arxiv.org/abs/1204.1737}{{\tt arXiv:1204.1737}}].

\bibitem{Buras:1990fn}
A.~J. Buras, M.~Jamin, and P.~H. Weisz, {\it {Leading and next-to-leading QCD
  corrections to $\varepsilon$ parameter and $B^0-\bar{B}^0$ mixing in the
  presence of a heavy top quark}},  {\em Nucl.~Phys.} {\bf B347} (1990)
  491--536.

\bibitem{Gambino:1998rt}
P.~Gambino, A.~Kwiatkowski, and N.~Pott, {\it {Electroweak effects in the $B^0
  - \bar B^0$ mixing}},  {\em Nucl. Phys.} {\bf B544} (1999) 532--556,
  [\href{http://arxiv.org/abs/hep-ph/9810400}{{\tt hep-ph/9810400}}].

\bibitem{Awramik:2003rn}
M.~Awramik, M.~Czakon, A.~Freitas, and G.~Weiglein, {\it {Precise prediction
  for the $W$ boson mass in the standard model}},  {\em Phys. Rev. D} {\bf 69}
  (2004) 053006, [\href{http://arxiv.org/abs/hep-ph/0311148}{{\tt
  hep-ph/0311148}}].

\bibitem{Zyla:2020zbs}
{\bf Particle Data Group} Collaboration, P.~A. Zyla et~al., {\it {Review of
  Particle Physics}},  {\em PTEP} {\bf 2020} (2020), no.~8 083C01.

\bibitem{Tanabashi:2018oca}
{\bf Particle Data Group} Collaboration, M.~Tanabashi et~al., {\it {Review of
  Particle Physics}},  {\em Phys. Rev. D} {\bf 98} (2018), no.~3 030001.

\bibitem{Aoki:2019cca}
{\bf Flavour Lattice Averaging Group} Collaboration, S.~Aoki et~al., {\it {FLAG
  Review 2019: Flavour Lattice Averaging Group (FLAG)}},  {\em Eur. Phys. J. C}
  {\bf 80} (2020), no.~2 113, [\href{http://arxiv.org/abs/1902.08191}{{\tt
  arXiv:1902.08191}}].

\bibitem{DiLuzio:2019jyq}
L.~Di~Luzio, M.~Kirk, A.~Lenz, and T.~Rauh, {\it {$\Delta M_s$ theory precision
  confronts flavour anomalies}},  {\em JHEP} {\bf 12} (2019) 009,
  [\href{http://arxiv.org/abs/1909.11087}{{\tt arXiv:1909.11087}}].

\bibitem{Beneke:2020fot}
M.~Beneke, C.~Bobeth, and Y.-M. Wang, {\it {$B_{d,s}\to\gamma\ell\bar{\ell}$
  decay with an energetic photon}},  {\em JHEP} {\bf 12} (2020) 148,
  [\href{http://arxiv.org/abs/2008.12494}{{\tt arXiv:2008.12494}}].

\bibitem{Bazavov:2017lyh}
A.~Bazavov et~al., {\it {$B$- and $D$-meson leptonic decay constants from
  four-flavor lattice QCD}},  \href{http://arxiv.org/abs/1712.09262}{{\tt
  arXiv:1712.09262}}.

\bibitem{Bussone:2016iua}
{\bf ETM} Collaboration, A.~Bussone et~al., {\it {Mass of the b quark and B
  -meson decay constants from $N_f=2+1+1$ twisted-mass lattice QCD}},  {\em
  Phys. Rev. D} {\bf 93} (2016), no.~11 114505,
  [\href{http://arxiv.org/abs/1603.04306}{{\tt arXiv:1603.04306}}].

\bibitem{Dowdall:2013tga}
{\bf HPQCD} Collaboration, R.~J. Dowdall, C.~T.~H. Davies, R.~R. Horgan, C.~J.
  Monahan, and J.~Shigemitsu, {\it {B-Meson Decay Constants from Improved
  Lattice Nonrelativistic QCD with Physical u, d, s, and c Quarks}},  {\em
  Phys. Rev. Lett.} {\bf 110} (2013), no.~22 222003,
  [\href{http://arxiv.org/abs/1302.2644}{{\tt arXiv:1302.2644}}].

\bibitem{Hughes:2017spc}
C.~Hughes, C.~T.~H. Davies, and C.~J. Monahan, {\it {New methods for $B$ meson
  decay constants and form factors from lattice NRQCD}},  {\em Phys. Rev. D}
  {\bf 97} (2018), no.~5 054509, [\href{http://arxiv.org/abs/1711.09981}{{\tt
  arXiv:1711.09981}}].

\bibitem{Bazavov:2016nty}
{\bf Fermilab Lattice, MILC} Collaboration, A.~Bazavov et~al., {\it
  {$B^0_{(s)}$-mixing matrix elements from lattice QCD for the Standard Model
  and beyond}},  {\em Phys. Rev.} {\bf D93} (2016), no.~11 113016,
  [\href{http://arxiv.org/abs/1602.03560}{{\tt arXiv:1602.03560}}].

\bibitem{Gamiz:2009ku}
{\bf HPQCD} Collaboration, E.~Gamiz, C.~T. Davies, G.~P. Lepage, J.~Shigemitsu,
  and M.~Wingate, {\it {Neutral $B$ Meson Mixing in Unquenched Lattice QCD}},
  {\em Phys.~Rev.} {\bf D80} (2009) 014503,
  [\href{http://arxiv.org/abs/0902.1815}{{\tt arXiv:0902.1815}}].

\bibitem{Aoki:2014nga}
Y.~Aoki, T.~Ishikawa, T.~Izubuchi, C.~Lehner, and A.~Soni, {\it {Neutral $B$
  meson mixings and $B$ meson decay constants with static heavy and domain-wall
  light quarks}},  {\em Phys. Rev. D} {\bf 91} (2015), no.~11 114505,
  [\href{http://arxiv.org/abs/1406.6192}{{\tt arXiv:1406.6192}}].

\bibitem{Dowdall:2019bea}
R.~J. Dowdall, C.~T.~H. Davies, R.~R. Horgan, G.~P. Lepage, C.~J. Monahan,
  J.~Shigemitsu, and M.~Wingate, {\it {Neutral $B$-meson mixing from full
  lattice QCD at the physical point}},  {\em Phys. Rev. D} {\bf 100} (2019),
  no.~9 094508, [\href{http://arxiv.org/abs/1907.01025}{{\tt
  arXiv:1907.01025}}].

\bibitem{King:2019lal}
D.~King, A.~Lenz, and T.~Rauh, {\it {$B_s$ mixing observables and
  $|V_{td}/V_{ts}|$ from sum rules}},  {\em JHEP} {\bf 05} (2019) 034,
  [\href{http://arxiv.org/abs/1904.00940}{{\tt arXiv:1904.00940}}].

\bibitem{Lenz:2019lvd}
A.~Lenz and G.~Tetlalmatzi-Xolocotzi, {\it {Model-independent bounds on new
  physics effects in non-leptonic tree-level decays of B-mesons}},  {\em JHEP}
  {\bf 07} (2020) 177, [\href{http://arxiv.org/abs/1912.07621}{{\tt
  arXiv:1912.07621}}].

\bibitem{Boyle:2018knm}
{\bf RBC/UKQCD} Collaboration, P.~A. Boyle, L.~Del~Debbio, N.~Garron,
  A.~Juttner, A.~Soni, J.~T. Tsang, and O.~Witzel, {\it {SU(3)-breaking ratios
  for $D_{(s)}$ and $B_{(s)}$ mesons}},
  \href{http://arxiv.org/abs/1812.08791}{{\tt arXiv:1812.08791}}.

\bibitem{Bediaga:2018lhg}
{\bf LHCb} Collaboration, R.~Aaij et~al., {\it {Physics case for an LHCb
  Upgrade II - Opportunities in flavour physics, and beyond, in the HL-LHC
  era}},  \href{http://arxiv.org/abs/1808.08865}{{\tt arXiv:1808.08865}}.

\bibitem{Buras:2005gr}
A.~J. Buras, M.~Gorbahn, U.~Haisch, and U.~Nierste, {\it {The rare decay $K^+
  \to \pi^+ \nu \bar\nu$ at the next-to-next-to-leading order in QCD}},  {\em
  Phys.~Rev.~Lett.} {\bf 95} (2005) 261805,
  [\href{http://arxiv.org/abs/hep-ph/0508165}{{\tt hep-ph/0508165}}].

\bibitem{Buras:2006gb}
A.~J. Buras, M.~Gorbahn, U.~Haisch, and U.~Nierste, {\it {Charm quark
  contribution to $K^+ \to \pi^+ \nu \bar\nu$ at next-to-next-to-leading
  order}},  {\em JHEP} {\bf 11} (2006) 002,
  [\href{http://arxiv.org/abs/hep-ph/0603079}{{\tt hep-ph/0603079}}].

\bibitem{Brod:2010hi}
J.~Brod, M.~Gorbahn, and E.~Stamou, {\it {Two-Loop Electroweak Corrections for
  the $K \to \pi \nu \bar{\nu}$ Decays}},  {\em Phys.~Rev.} {\bf D83} (2011)
  034030, [\href{http://arxiv.org/abs/1009.0947}{{\tt arXiv:1009.0947}}].

\bibitem{Buchalla:1996fp}
G.~Buchalla and A.~J. Buras, {\it {$K \to\pi\nu\bar\nu$ and high precision
  determinations of the CKM matrix}},  {\em Phys.~Rev.} {\bf D54} (1996)
  6782--6789, [\href{http://arxiv.org/abs/hep-ph/9607447}{{\tt
  hep-ph/9607447}}].

\bibitem{Giudice:2015toa}
G.~F. Giudice, P.~Paradisi, and A.~Strumia, {\it {Indirect determinations of
  the top quark mass}},  {\em JHEP} {\bf 11} (2015) 192,
  [\href{http://arxiv.org/abs/1508.05332}{{\tt arXiv:1508.05332}}].

\bibitem{Buttazzo:2013uya}
D.~Buttazzo, G.~Degrassi, P.~P. Giardino, G.~F. Giudice, F.~Sala, et~al., {\it
  {Investigating the near-criticality of the Higgs boson}},  {\em JHEP} {\bf
  1312} (2013) 089, [\href{http://arxiv.org/abs/1307.3536}{{\tt
  arXiv:1307.3536}}].

\end{thebibliography}\endgroup

\end{document}